\documentclass{jpp}
\usepackage{graphicx}
\usepackage{epstopdf, epsfig}
\usepackage{hyperref}

\usepackage{delarray}
\usepackage{here}
\usepackage{amsmath,amssymb}
\usepackage{inputenc}

 \newcommand{\be}{\begin{displaymath}}
 \newcommand{\bn}{\begin{equation}}
 \newcommand{\bea}{\begin{eqnarray*}}
 \newcommand{\eea}{\end{eqnarray*}}
 \newcommand{\en}{\end{equation}}
 \newcommand{\ee}{\end{displaymath}}
 \newcommand{\lang}{\left\langle}
 \newcommand{\rang}{\right\rangle}

\usepackage{xcolor}

\shorttitle{Energetic bounds on gyrokinetic instabilities.  Part I.}
\shortauthor{P. Helander and G.G. Plunk}

\title{Energetic bounds on gyrokinetic instabilities. Part I: Fundamentals}

\author{P. Helander\aff{1}
  \corresp{\email{per.helander@ipp.mpg.de}} 
	\and G. G. Plunk\aff{1}}

\affiliation{\aff{1}Max Planck Institute for Plasma Physics, Greifswald, Germany}

\begin{document}

\maketitle

\begin{abstract}
Upper bounds on the growth of free energy in gyrokinetics are derived. These bounds apply to all local gyrokinetic instabilities in the geometry of a flux tube, i.e. a slender volume of plasma aligned with the magnetic field, regardless of the geometry of field, the number of particle species, or collisions. The results apply both to linear instabilities and to the nonlinear growth of finite-amplitude fluctuations.
\end{abstract}

 \section{Introduction}

 For the last six and a half decades, an enormous effort has been devoted to the study of microinstabilities in magnetically confined plasmas. Mathematically, such instabilities can be desribed by the Boltzmann equation for the plasma particles coupled to Maxwell's equations for the electric and magnetic fields, but it is often sufficient to consider the somewhat simpler gyrokinetic system of equations \citep{Taylor-1968,Rutherford-1968-a,Antonsen-1980,Catto-1978-a,Catto-1981,Frieman-1982,Brizard-2007,Krommes-2012,Catto-2019}. These equations apply if the instability wavelength perpendicular to the magnetic field is comparable to the ion or electron gyroradius but the wavelength is much longer in the direction along the field, which is normally the case for the most important microinstabilities and turbulence afflicting magnetised plasmas in the laboratory. Gyrokinetics also finds fruitful application in other parts of plasma physics, such as astrophysics \citep{Schekochihin-2009}, and has been the subject of thousands of publications. Several millions of lines of computer code has been written for the purpose of numerically simulating gyrokinetic instabilities and turbulence \citep{Kotschenreuther-1995,Garbet-2010}.

As a result of this effort, a great deal of knowledge about various microinstabilities has accumulated. Ion- and electron-temperature-gradient-driven modes, trapped electron modes, kinetic ballooning modes and microtearing modes have, for instance, been found to be unstable and cause turbulence in tokamaks, stellarators, and other fusion devices. However, a basic problem is that these and other instabilities tend to be sensitive to assumptions made about plasma parameters and the magnetic-field geometry. A cylindrical plasma does not have the same stability properties as a plasma slab, toroidal plasmas are different from cylindrical ones, and tokamaks and stellarators are also substantially different. As a result, little is known {\em in general} about gyrokinetic microinstaiblities, despite the great effort devoted to their study. 

In a recent publication \citep{Helander-Plunk-2021}, universal upper bounds on the growth rates of local gyrokinetic instabilities could nevertheless be derived in such a way that the results hold in any low-beta plasma, regardless of the magnetic geometry, number of particle species, and collisions. The reason why these bounds are so general is they result from thermodynamic considerations. It is the budget of the Helmholtz free energy that constrains all instability growth rates to lie below the bounds in question. In the present paper, we provide more mathematical details of this calculation and extend it by showing how the bounds can be sharpened. In particular, we calculate the lowest possible bound on the growth rate that can be obtained from the free-energy budget of a plasma with ``adiabatic'' electrons and a single kinetic ion species. In subsequent publications, such rates of ``optimal growth'' will be derived in more complex cases that include both electrostatic and magnetic fluctuations. We will also show how the bounds can be lowered by simulataneously considering the budget of free energy and electrostatic energy, and compare them with gyrokinetic simulations. The present paper serves as an introduction to this series of publications. 

\section{Gyrokinetic system of equations}

The mathematical setting of our considerations is that of local gyrokinetics. The distribution function of each species $a$ is written as \citep{Catto-1978-a}
	$$ f_a({\bf r}, E_a, \mu_a, t) = F_{a0} ({\psi}, E_a) \left( 1 - \frac{e_a \delta \phi({\bf r})}{T_a} \right) 
	+ g_a({\bf R}, E_a, \mu_a, t), $$
where ${\bf r}$ denotes the particle position and ${\bf R} = {\bf r} - {\bf b} \times {\bf v} / \Omega_a$ the gyrocentre position. Here, the magnetic field has been written as ${\bf B} = B {\bf b} = \nabla \psi \times \nabla \alpha$ in terms of Clebsch coordinates $(\psi,\alpha)$. If the magnetic field lines trace out toroidal surfaces, as in tokamaks and stellarators, a ballooning transform is necessary unless all field lines close on themselves. The gyrofrequency is $\Omega_a = e_a B / m_a$, where $m_a$ denotes mass and $e_a$ charge. The equilibrium distribution function is taken to be Maxwellian, with density $n_a(\psi)$ and temperature $T_a(\psi)$ constant on magnetic surfaces, and no mean flow velocity. The particle velocity is denoted ${\bf v} = v_\| {\bf b} + {\bf v}_\perp$, the unperturbed energy by $E_a = m_a v^2 / 2 + e_a \Phi(\psi)$, and the magnetic moment $\mu_a = m_a v_\perp^2 / (2B)$ is a lowest-order constant of the motion. The geometry is taken to be that of a ``flux tube'', i.e. a slender volume of plasma aligned with the magnetic field, with a rectangular cross section in the $(\psi,\alpha)$-plane. Periodic boundary conditions on the fluctuations will be applied in this plane, so that all perturbations can be Fourier decomposed. For instance, the electrostatic potential $\delta \phi$ fluctuations are
	$$ \delta \phi(\psi,\alpha,l) = \sum_{\bf k} \delta \phi_{\bf k} (l) e^{i(k_\psi \psi + k_\alpha \alpha)}, $$
where ${\bf k} = {\bf k}_\perp = k_\psi \nabla \psi + k_\alpha \nabla \alpha$ with $k_\psi$ and $k_\alpha$independent of the arc length $l$ along the magnetic field. The Fourier coefficients must satisfy $\delta \phi^\ast_{\bf k} = \delta \phi_{-\bf k}$ in order that the potential be real. 

The ``non-adiabatic'' part of the distribution function $g_a$ evolves according to the non-linear gyrokinetic equation \citep{Frieman-1982}
	$$  \frac{\partial g_{a,{\bf k}}}{\partial t} +
	v_{\|} \frac{\partial g_{a,{\bf k}}}{\partial l} + i \omega_{da} g_{a, {\bf k}} 
	+ \frac{1}{B^2} \sum_{{\bf k}'} {\bf B} \cdot ({\bf k} \times {\bf k}')  
	\bar{\chi}_{a, {\bf k}'} g_{a, {\bf k} - {\bf k}'}
	$$
	\begin{equation} = \sum_b \left[ C_{ab}(g_{a, \bf k},F_{b0}) + C_{ab}(F_{a0},g_{b, \bf k}) \right] 
	+ \frac{e_a F_{a0}}{T_a} 
	\left( \frac{\partial}{\partial t} + i \omega_{\ast a}^T \right) \bar{\chi}_{a, \bf k} 
	\label{gk},
	\end{equation}
where $\omega_d = {\bf k} \cdot {\bf v}_d$ denotes the drift frequency (with ${\bf v}_d$ being the unperturbed drift velocity), 
	$$ \omega_{\ast a} = \frac{k_\alpha T_a}{e_a} \frac{d \ln n_a}{d \psi}, $$
	$$ \omega_{\ast a}^T = \omega_{\ast a} 
	\left[1 + \eta_a \left( \frac{m_a v^2}{2 T_a} - \frac{3}{2} \right)\right], $$
	$$ \bar{\chi}_{a \bf k} 
	= J_0 \left( \frac{k_\perp v_\perp}{\Omega_a} \right)\left( \delta \phi_{{\bf k}} - v_\| \delta A_{\| {\bf k}} \right)
	+ J_1 \left( \frac{k_\perp v_\perp}{\Omega_a} \right) \frac{v_\perp}{k_\perp} \delta B_{\| \bf k}, $$
and $J_0$ and $J_1$ are Bessel functions. The gyro-averaged and linearised collision operator between species $a$ and $b$ is denoted by $C_{ab}$, and the field perturbations are given by 
	\bn \sum_a \lambda_a \delta \phi_{\bf k} = \sum_a e_a \int g_{a, {\bf k}} J_{0a} d^3v, 
	\label{field1}
	\en
	\bn \delta A_{\| {\bf k}} = \frac{\mu_0}{k_\perp^2} \sum_a e_a \int v_\| g_{a, {\bf k}} J_{0a} d^3v, 
	\label{field2}
	\en
	\bn \delta B_{\| {\bf k}} = - \frac{\mu_0}{k_\perp} \sum_a e_a \int v_\perp g_{a, {\bf k}} J_{1a} d^3v.
	\label{field3}
	\en
Here and in the following, we write $ \lambda_a = {n_a e_a^2}/{T_a}$ and $J_{na} = J_n(k_\perp v_\perp / \Omega_a)$. Equation\,(\ref{field1}) expresses quasineutrality, Eq.\,(\ref{field2}) Amp{\`e}re's law, and Eq.\,(\ref{field3}) the condition that the sum of the thermal pressure and the magnetic pressure should be constant on the short length scale of the fluctuations. The space volume element in velocity space is
	$$ d^3v = 2 \pi v_\perp dv_\perp v_\| = \sum_\sigma \frac{2 \pi B dE_a d\mu_a}{m_a^2 | v_\| |}, $$
where the sum is taken over both values of $\sigma = v_\| / | v_\| | = \pm 1$. 

As we shall see below, it is advantageous to introduce the function
	\bn \delta F_{a, \bf k} = g_{a, \bf k} - \frac{e_a J_{0a} \delta \phi_{\bf k}}{T_a}, 
	\label{delta F}
	\en
where all quantities are evaluated at the gyro-centre position $\bf R$. The quasineutrality condition then becomes
	\bn \sum_a \lambda_a \left[1 - \Gamma_0(b_a) \right] \delta \phi_{\bf k} 
	= \sum_a e_a \int \delta F_{a, {\bf k}} J_{0a} d^3v, 
	\label{field4}
	\en
where $\Gamma_0(x) = I_0(x) e^{-x}$ and $b_a = k_\perp^2 \rho_a^2 = k_\perp^2 T_a / (m_a \Omega_a^2)$. In the following, we shall sometimes write $\Gamma_{0a}$ instead of $\Gamma_0(b_a)$.

\section{Helmholtz free energy}

The budget of Helmholtz free energy has been considered by several authors, e.g. \cite{Krommes-1993,Brizard-1994,Sugama-1996,Garbet-2005,Schekochihin-2009,Banon-2011,Hatch-2016,Stoltzfus-Dueck-2017}, and is obtained by multiplying the gyrokinetic equation (\ref{gk}) by $T_a g_a^\ast / F_{a0}$, taking the real part, summing over all species and wave numbers, integrating over velocity space, and finally taking an average over the volume of the flux tube, which we denote by angular brackets, 
	$$ \lang \cdots \rang = \lim_{L\rightarrow \infty} \int_{-L}^L (\cdots ) \frac{dl}{B} \bigg\slash
	\int_{-L}^L \frac{dl}{B}.
	$$
We note that the average could also be defined keeping $L$ finite, {\em e.g.} for periodic systems, without affecting what follows.  In order for the integral to converge, we require that the functions $\bar{\chi}_{\bf k}(l)$ should be bounded. On the left-hand side of Eq.\,(\ref{gk}), this operation, 
	$$ {\rm Re} \; \sum_{a,{\bf k}} T_a \lang \int \left( \cdots \right) \frac{g_{a, \bf k}^\ast}{F_{a0}} d^3v \rang, $$
annihilates the second term since
	$$ {\rm Re} \; \lang g_{a, {\bf k}}^\ast v_\| \frac{\partial g_{a, {\bf k}}}{\partial l} d^3v \rang
	\propto 
	\lim_{L\rightarrow \infty}\int_{-L}^L \frac{\partial |g_{a, {\bf k}}|^2}{\partial l} dl \bigg\slash \int_{-L}^L \frac{dl}{B} = 0. $$
where we have have used that $d^3v \propto B/|v_\||$, and assumed that $|g_{a, {\bf k}}|^2$ remains bounded as $\ell \rightarrow \infty$, so that ratio goes to zero.\footnote{For finite systems, Dirichlet boundary conditions, $g_{a, {\bf k}}(\pm L) = 0$ (as used in gyrokinetic simulations), or periodic boundary conditions, $g_{a, {\bf k}}(L) = g_{a, {\bf k}}|_(-L)$, work equally well here.}  The operation also eliminates the third term since $\omega_{da}$ is real and the fourth term since
	$$ {\rm Re} ({\bf k} \times {\bf k}')  
	g^\ast_{a,\bf k} \bar{\chi}_{a, {\bf k}'} g_{a,{\bf k} - {\bf k}'} 
	= \frac{1}{2} \left[ ({\bf k} \times {\bf k}') 
	\left( g^\ast_{a,\bf k} \bar{\chi}_{a, {\bf k}'} g_{a,{\bf k} - {\bf k}'} + g_{a,\bf k} \bar{\chi}^\ast_{a, {\bf k}'} g^\ast_{a,{\bf k} - {\bf k}'} \right) \right]
	$$
	$$ = \frac{1}{2} \left[ ({\bf k} \times {\bf k}') 
	\left( g_{a,-\bf k} \bar{\chi}_{a, {\bf k}'} g_{a,{\bf k} - {\bf k}'} + g_{a,\bf k} \bar{\chi}_{a,- {\bf k}'} g_{a,{\bf k}' - {\bf k}} \right) \right]
	$$
vanishes upon summation over $\bf k$ and ${\bf k}'$. The remainder of the equation thus becomes
	$$ \frac{d}{dt} \sum_{a,{\bf k}}  T_a \lang \int \frac{|g_{a \bf k}|^2}{2 F_{a0}} d^3v \rang =  
	\sum_{\bf k} C({\bf k},t) 
	+ {\rm Re} \sum_{a,{\bf k}} 
	\lang \int g_{a,{\bf k}}^\ast \left( \frac{\partial}{\partial t} + i \omega_{\ast a}^T \right) \bar{\chi}_{a \bf k} d^3v \rang, $$
where 	
	\bn C({\bf k},t) = {\rm Re} \; \sum_{a,b} T_a \lang \int \frac{g_{a, \bf k}^\ast}{ F_{a0}} 
	\left[C_{ab}(g_{a, \bf k},F_{b0}) + C_{ab}(F_{a0},g_{b, \bf k}) \right] d^3v \rang \le 0 
	\label{H-theorem}
	\en
is negative or vanishes by Boltzmann's H-theorem. By using the field equations (\ref{field1})-(\ref{field3}), we find
	$$ \sum_{a} \int g_{a,{\bf k}}^\ast \frac{\partial \bar{\chi}_{a \bf k}}{\partial t}  d^3v 
	= \frac{1}{2} \frac{d}{dt} 
	\left( \sum_{a} \lambda_a |\delta \phi_{\bf k}|^2 - \frac{| \delta {\bf B}_{\bf k} |^2}{\mu_0}\right), $$
where $| \delta {\bf B}_{\bf k} |^2 = | k_\perp \delta A_{\|\bf k} |^2 + | \delta B_{\|\bf k} |^2$, and thus we obtain our key equation: 
	\bn \frac{d}{dt} \sum_{\bf k} H({\bf k},t) = 2 \sum_{\bf k} \left[C({\bf k},t) + D({\bf k},t)) \right], 
	\label{entropy balance}
	\en
where we have written
	\bn D({\bf k}, t) = {\rm Im} \; \sum_a  e_a \lang \int g_{a, \bf k} \omega_{\ast a}^T 
	\bar{\chi}^\ast_{a,\bf k} d^3v \rang, 
	\label{D1}
	\en
	$$ H({\bf k},t)  = \sum_a \lang  T_a \int \frac{|g_{a, \bf k}|^2}{F_{a0}} d^3v - \lambda_a |\delta \phi_{\bf k}|^2 \rang
	+ \lang \frac{| \delta {\bf B}_{\bf k} |^2}{\mu_0} \rang. $$

It is helpful to write $H$ in terms of $\delta F_a$, defined in Eq.\,(\ref{delta F}), instead of $g_a$:
	$$ H({\bf k},t)  = \sum_a \lang  T_a \int \frac{|\delta F_{a, \bf k}|^2}{F_{a0}} d^3v 
	+ \lambda_a (1 - \Gamma_{0a} ) | \delta \phi_{\bf k}|^2 \rang
	+ \lang \frac{| \delta {\bf B}_{\bf k} |^2}{\mu_0} \rang, $$
which makes it clear that $H$ can never be negative and only vanishes if all distribution-function perturbations $\delta F_a$ vanish everywhere in phase space. The first term in $H$ is recognised from the Gibbs entropy formula: if $F = F_0 + \delta F$, then to second order in $\delta F$
	$$ - \int F \ln F \; d^3v 
	= - \int \left[ F_0 \ln F_0  + \left( 1 + \ln F_0 \right) \delta F 
	+ \frac{\delta F^2}{2 F_0} \right] d^3v, $$
which motivates us to define
	$$ S_a({\bf k},t) = - \lang \int \frac{|\delta F_{a,  \bf k}|^2}{F_{a0}} d^3v \rang. $$
Furthermore, we write 
	$$ U({\bf k},t) = \lang \sum_{a} \lambda_a (1 - \Gamma_{0a} ) | \delta \phi_{\bf k}|^2 
	+ \frac{| \delta {\bf B} |^2}{\mu_0} \rang, $$
and note that, in the short-wavelength limit, $b_a = (k_\perp \rho_a)^2 \ll 1$, $\Gamma_0(b_a) = 1 - b_a + O(b_a^2)$, so that
	$$ U({\bf k},t) = \lang \sum_{a} \frac{m_a n_a k^2 | \delta \phi_{\bf k}|^2}{B^2} 
	+ \frac{| \delta {\bf B} |^2}{\mu_0} \rang, $$
where the first term represents the kinetic energy of ${\bf E} \times {\bf B}$ motion and the second term magnetic energy. We thus arrive at the formula
		$$ H({\bf k},t) = U ({\bf k},t) - T_a S_a({\bf k},t), $$	
with $U$ denoting the energy of the fluctuations and $S_a$ their entropy, suggesting that $H$ describes the Helmholtz free energy of the fluctuations and Eq.\,(\ref{entropy balance}) the budget of this energy. Indeed, on the right-hand side of this equation $C$ reflects the increase in entropy due to collisions, and $D$ can be written as
	$$ D({\bf k},t) = {\rm Re} \; \sum_a  T_a \left\langle \int g_a 
	\delta \dot {\bf R}_{a, \bf k}^\ast
	\cdot \nabla F_{a0} d^3v \; \right\rangle $$
	\bn = - \; \sum_a  
	\left ( T_a\Gamma_a \frac{d \ln p_a}{d \psi} + q_a \frac{d \ln T_a}{d \psi} \right). 
	\label{interpretation of D}
	\en
Here
		$$ \delta \dot {\bf R}_{a, \bf k} = \frac{i \bar{\chi}_{a, \bf k}{\bf b} \times {\bf k}}{B} $$
describes the gyro-centre velocity perturbation due to the fluctuations, and the radial particle and heat fluxes are
	$$ \Gamma_a({\bf k},t) =  {\rm Re}\lang \int \delta F_{a, \bf k} (\delta \dot {\bf R}_{a, \bf k}^\ast \cdot \nabla \psi)d^3v \rang, $$
	$$ q_a({\bf k},t) =  {\rm Re}\lang \int \delta F_{a, \bf k} \left( \frac{m_a v^2}{2T_a} - \frac{5 T_a}{2} \right)
	(\delta \dot {\bf R}_{a, \bf k}^\ast \cdot \nabla \psi)d^3v \rang. $$
The term in (\ref{interpretation of D}) involving $\Gamma_a$ is thus suggestive of the thermodynamic work performed by the particle flux against the pressure gradient, and the term involving $q_a$ relates to entropy production due to a heat flux down the temperature gradient.

Thanks to the nonlinear term in the gyrokinetic equation, free energy can be transferred between different wave numbers and be ``cascaded'' to small scales, where it is dissipated by collisions, much like kinetic energy in Navier-Stokes turbulence. The way in which this occurs and gives rise to a turbulent spectrum of fluctuations has been studied extensively in the literature \citep{Schekochihin-2009,Tatsuno-2009,Banon-2011,Stoltzfus-Dueck-2017}. We shall use the free-energy budget (\ref{entropy balance}) for a different purpose, namely, to derive rigorous upper bounds on linear and nonlinear growth rates. Outside the realm of gyrokinetics, this has earlier been accomplished for linear instabilites by Fowler and co-workers \citep{Fowler-1964,Fowler-1968,Brizard-1991}. 

\section{Cauchy-Schwarz inequalities}

For simplicity, we restrict our considerations to low-beta plasmas, where fluctuations in the magnetic-field strength can be neglected, $\delta B_\| = 0$. This approximation is common in the literature but will be removed in the next publication in this series of papers. 

Our basic mathematical tools are the triangle and Cauchy-Schwarz inequalities, which limit the amplitude of field fluctuations that are possible given a certain entropy budget. For instance, it follows from the field equation (\ref{field4}) that the electrostatic potential is bounded by
	$$ \sum_a \lambda_a \left(1 - \Gamma_{0a} \right) |\delta \phi_{\bf k}  | \le 
	\sum_a | e_a | \left( \int \frac{ |\delta F_{a,{\bf k}} |^2}{F_{a0}} d^3v \int F_{a0} J_{0a}^2 d^3v \right)^{1/2}. $$
Thus, if we measure the relative entropy perturbation at the scale $\bf k$ of each species $a$ by the dimensionless quantity 
	$$ s_a({\bf k},t) = \frac{1}{n_a} \int \frac{|\delta F_{a \bf k} |^2}{F_{a0}} d^3v, $$
then it follows that the electrostatic potential is subject to the bound
	\bn \sum_a \lambda_a \left(1 - \Gamma_{0a} \right) |\delta \phi_{\bf k}  | \le \sum_a n_a |e_a| \sqrt{ \Gamma_{0a} s_a}.
	\label{bound on phi}
	\en
Analogously, it follows from Amper{\`e}'s law (\ref{field2}) that the magnetic potential is limited by
		$$ | \delta A_{\| \bf k} | \le \frac{\mu_0 | e_a |}{k_\perp^2} \left( \int \frac{ |\delta F_{a,{\bf k}} |^2}{F_{a0}} d^3v \int v_\|^2 F_{a0} J_{0a}^2 d^3v \right)^{1/2}, $$
i.e., 
	\bn \frac{k_\perp | \delta A_{\| \bf k} |}{B} \le \sum_a \frac{\beta_a}{2 k_\perp \rho_a} \sqrt{\Gamma_{0a} s_a} \simeq \frac{\beta_e}{2 k_\perp \rho_e} \sqrt{\Gamma_{0e} s_e},
	\label{bound on A}
	\en
where $\beta_a(l) = 2 \mu_0 n_a T_a / B^2$. In the last, approximate equality, we have recognised the fact that the sum is usually dominated by the contribution from the electrons thanks to their small gyroradius. 

We can also apply the triangle and Cauchy-Schwarz inequalities to the free-energy production rate (\ref{D1}): 
	$$ D({\bf k},t) \le \sum_a |e_a| |n_a s_a|^{1/2} 
	\lang \int F_{a0} (\omega_{\ast a}^T)^2 J_0^2 \left(|\delta \phi_{\bf k}  |^2 + v_\|^2 |\delta A_{\| \bf k}|^2 \right) d^3v \rang^{1/2}  $$
	\bn = \sum_a n_a |e_a \omega_{\ast a}| |s_a|^{1/2} 
	\lang M(\eta_a, b_a) |\delta \phi_{\bf k}  |^2 +
	N(\eta_a, b_a) \frac{T_a |\delta A_{\| \bf k}|^2}{m_a}  \rang^{1/2}, 
	\label{D}
	\en
where the functions
	$$ M(\eta_a ,b_a) = \frac{1}{n_a} \int \left[1 + \eta_a \left( \frac{m_a v^2}{2 T_a} - \frac{3}{2} \right) \right]^2 F_{a0} J_{0a}^2 d^3v, $$
	$$ N(\eta_a ,b_a) = \frac{1}{n_a} \int \frac{m_a v_\|^2}{T_a} \left[1 + \eta_a \left( \frac{m_a v^2}{2 T_a} - \frac{3}{2} \right) \right]^2 F_{a0} J_{0a}^2 d^3v $$
can be expressed in terms of modified Bessel functions as
	$$ M(\eta,b) = \left( 1 + \frac{3 \eta^2}{2} - 2\eta(1+ \eta) b + 2 \eta^2 b^2 \right) \Gamma_0(b)
	+ \eta b\left( 2 + \eta - 2 \eta b \right) \Gamma_1(b), $$
	$$ N(\eta,b) = \left( 1 + 2 \eta + \frac{7 \eta^2}{2} - 2\eta(1+ 2\eta) b + 2 \eta^2 b^2 \right) \Gamma_0(b)
	+ \eta b\left( 2 + 3 \eta - 2 \eta b \right) \Gamma_1(b). $$	
In the limits of very small and very large wavelength, respectively, the asymptotic forms of the functions $\Gamma_0(b)$ and $\Gamma_1(b)$ are

\begin{eqnarray}
  \Gamma_0(b) \simeq \left\{
    \begin{array}{cl}
      1 -b , & b \rightarrow 0, \\
      \frac{1}{\sqrt{2 \pi b}} \left( 1 + \frac{1}{8b} + \frac{9}{128 b^2} \right), & b \rightarrow \infty,
    \end{array} \right.\\
  \Gamma_1(b) \simeq \left\{
    \begin{array}{cl}
      b , & b \rightarrow 0, \\
      \frac{1}{\sqrt{2 \pi b}} \left( 1 -  \frac{3}{8b} - \frac{15}{128 b^2} \right), & b \rightarrow \infty,
    \end{array} \right.
\end{eqnarray}
and those for $M(\eta,b)$ and $N(\eta,b)$
\begin{eqnarray}
  M(\eta,b) \simeq \left\{
    \begin{array}{cl}
      1 + \frac{3 \eta^2}{2}, & b \rightarrow 0 \\
      \frac{1-\eta + \frac{5 \eta^2}{4}}{\sqrt{2 \pi b}}, & b \rightarrow \infty,
    \end{array} \right.\label{M}\\
  N(\eta,b) \simeq \left\{
    \begin{array}{cl}
      1 + 2 \eta + \frac{7 \eta^2}{2}, & b \rightarrow 0 \\
      \frac{1+\eta + \frac{9 \eta^2}{4}}{\sqrt{2 \pi b}}, & b \rightarrow \infty.\label{N}
    \end{array} \right.
\end{eqnarray}

\section{Upper bounds on linear growth rates}

In this section, we temporarily consider linear instabilities and thus focus on a single pair of wave numbers $(k_\psi,k_\alpha)$. Thanks to Boltzmann's $H$-theorem, the quantity $C({\bf k},t)$ is always negative and the relation (\ref{entropy balance}) thus implies an upper bound on the linear growth rate
	\bn \gamma({\bf k}) \le \frac{D({\bf k},t)}{H({\bf k},t)}. 
	\label{bound on linear growth}
	\en
Since we have already bounded $D$ from above, we merely need to find a suitable bound on 
	\bn H({\bf k},t) = \sum_a \lang n_a T_a s_a + \lambda_a \left( 1 - \Gamma_{0a} \right) |\delta \phi_{\bf k}|^2 \rang + \lang \frac{| k_\perp \delta A_{\| \bf k} |^2}{\mu_0} \rang
	\label{H}
	\en
from below to derive an upper bound on $\gamma({\bf k})$. Some care is needed to construct reasonably tight bounds, but all results are largely independent of the geometry of the magnetic field since the second and third terms from Eq.~(\ref{gk}) do not contribute to the free-energy balance equation (\ref{entropy balance}). The bound (\ref{bound on linear growth}) therefore only depends on the magnetic geometry through the two quantities $B(l)$ and $ k_\perp(l) = \left| k_\psi \bnabla \psi+ k_\alpha \bnabla \alpha \right|$.

\subsection{Adiabatic electrons}

We begin by considering the simplest case of a hydrogen plasma with a Boltzmann-distributed, or so-called ``adiabatic'', electron response, where $g_e$ is taken to vanish. This is the traditionally simplest gyrokinetic model of ion-temperature-gradient (ITG) and trapped-ion instabilities, which account for a substantial fraction of the turbulence and transport in tokamaks and stellarators, and therefore has been the subject of hundreds, if not thousands, of publications. Since $g_e$ vanishes and there are no magnetic fluctuations, the free energy becomes
	$$ {H} = {n T_i}\lang s_i + \left(1 +\tau - \Gamma_{0i} \right) \left| \frac{e \delta \phi_{\bf k}}{T_i} \right|^2 \rang. $$
where $n = n_i = n_e$ and $\tau = T_i / T_e$. Furthermore, the quasineutrality condition (\ref{field4}) reduces to 
	\bn \left(1 +\tau - \Gamma_{0i} \right) \frac{e \delta \phi_{\bf k}}{T_i} = \frac{1}{n} \int \delta F_i J_{0i} d^3v, 
	\label{QN ae}
	\en
and the bound (\ref{bound on phi}) is thus replaced by the more stringent condition
	$$ \left(1 +\tau - \Gamma_{0i} \right) \frac{e |\delta \phi_{\bf k}|}{T_i} \le \sqrt{ \Gamma_{0i} s_{i}}.$$
Thanks to this inequality, the free energy satisfies
	$$ H \ge \lang \frac{1+\tau}{\Gamma_{0i}} \left(1 +\tau - \Gamma_{0i} \right) \left| \frac{e \delta \phi_{\bf k}}{T_i} \right|^2 \rang.  $$

The free-energy production term can be simplified somewhat since the quasineutrality condition (\ref{QN ae}) in the case of adiabatic electrons implies that there is no particle flux and $D$ thus becomes
	$$ D({\bf k},t) = {\rm Im} \;  \eta_i \omega_{\ast i} \lang e \delta \phi_{\bf k}^\ast\int g_{i\bf k} \left( \frac{m_i v^2}{2 T_i} - \frac{3}{2} \right) J_{0i} d^3v \rang. $$
As a result,  in the inequality (\ref{D}), the function $M(\eta,b)$ can be replaced by 
	$$ \tilde M(\eta,b) = \eta^2 \left[ \left(\frac{3 }{2} - 2b + 2 b^2 \right) \Gamma_0(b)
	+ b\left(1 - 2  b \right) \Gamma_1(b) \right], $$
and the bound (\ref{bound on linear growth}) becomes
	$$ \frac{\gamma}{\omega_{\ast i}} \le 
	\frac{\lang \tilde M(\eta_i, b_i) |\delta \phi_{\bf k} |^2 \rang^{1/2}}{ \lang (1+\tau) [(1+\tau) \Gamma_{0i}^{-1} - 1]|\delta \phi_{\bf k} |^2 \rang^{1/2}}. $$
The right-hand side is maximised by choosing $|\delta \phi_{\bf k}(l)|^2 = \delta(l-l_0)$, where $l_0$ is position along the field line where $b_i(l) = k_\perp^2 \rho_i^2 \propto (k_\perp/B)^2$ is minimised. We thus obtain
	\bn \frac{\gamma}{\omega_{\ast i}} \le \sqrt{ \frac{ \tilde M(\eta_i, b_{\rm min})}
	{(1+\tau) \left[ (1 + \tau) \Gamma_0^{-1} (b_{\rm min}) - 1 \right]}}, 
	\label{ae}
	\en
where $b_{\rm min} = b_i(l_0)$. The result is plotted in Fig.~1. Note that all dependence on the geometry of the magnetic field has disappeared: our limit on the growth rate is spatially local in nature and only depends on the minimum value of $k_\perp \rho_i$.

\begin{figure}
  \centerline{\includegraphics[width=\textwidth]{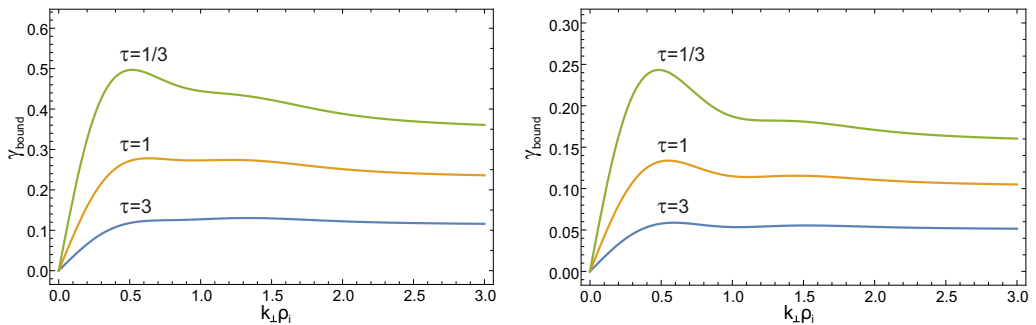}}
  \caption{The left panel shows the upper bound (\ref{ae}) on the growth rate normalised to $\eta_i \omega_{\ast i} / (k_\perp \rho_i)$ of gyrokinetic instabilities for $k_\psi = 0$ and three different values of $\tau = T_i/T_e$ in a hydrogen plasma with adiabatic electrons as a function of the smallest value of $k_\perp \rho_i$ along the magnetic field. The best possible bound (\ref{optimal ae bound}) for free-energy growth is about a factor of 2 lower and is plotted in the right panel. }
\label{fig:figure1}
\end{figure}

This bound, which applies to all local gyrokinetic instabilities in a plasma with adiabatic electrons, is not optimal and can be improved by a factor of about 2, as we shall see in the next section. Nevertheless, it displays scalings that have been seen in many publications and numerical simulations over the years. For long wavelengths, $b_i \rightarrow 0$, it reduces to 
	$$ \gamma \le |\eta_i \omega_{\ast i}| \sqrt{ \frac{3}{2\tau(1+\tau)}}. $$
Note that all dependence on the magnetic geometry has disappeared, and since $\omega_{\ast i} \propto k_\alpha$ the growth rate is proportional to $k_\alpha$ in this limit.  For short wavelengths, $k_\perp \rho_i \gg 1$, the bound remains finite, 
	$$ \gamma \le  \frac{| \eta_i \omega_{\ast i} |}{1+\tau} 
	\sqrt{ \frac{ 5}{8 \pi b_{\rm min}} }, $$	
since 
	$$ b_{\rm min}  = \min_l \left[ \left( k_\psi^2 |\bnabla \psi |^2 + 2 k_\psi k_\alpha \bnabla \psi \cdot \bnabla \alpha + k_\alpha^2 |\bnabla \alpha |^2 \right) 
	\frac{T_i}{m_i \Omega_i^2} \right] $$ 
is a positive-definite quadratic form in $k_\psi$ and $k_\alpha$. Indeed, $\gamma(k_\psi,k_\alpha)$ approaches a finite constant in the limit $k_\alpha \rightarrow \infty$ and vanishes if $k_\psi \rightarrow \infty$ at fixed $k_\alpha$. Moreover, at constant ion temperature, the bound (\ref{ae}) increases with the electron temperature through the scaling with $\tau$, which is a well-known feature of numerical simulations and analytical dispersion relations in explicitly tractable limits \citep{Bigliari-1989,Romanelli-1989,Plunk-2014,Zocco-2018}. This unfortunate scaling is thought to degrade energy confinement in electron-heated tokamaks and stellarators. 

\subsection{Electromagnetic instabilities}

We now turn to the more general case of an arbitrary number of kinetic species, but still restrict our attention to instabilities with $\delta B_\| = 0$. No attempt will be made to make the bound as low as possible. Our main concern is to show that an upper bound exists and that it is itself bounded as a function of $\bf k$, so that there is a universal upper bound on the growth rate at any wavelength. This will be of crucial importance when we consider nonlinear growth in a subsequent section. In the next publication of this series, we show how to extend the calculation to include fluctuations of the magnetic field strength and how to compute the lowest possible bounds in this context. 

We begin by seeking lower bounds on $H$ under the constraints (\ref{bound on phi}) and (\ref{bound on A}), which lead us to a simple quadratic minimisation problem treated in the Appendix. In terms of the notation used there, we first choose $x_a = \sqrt{s_a}$, $p_a = n_a |e_a| \sqrt{\Gamma_{0a}}$, $q_a = n_a T_a$ and
	$$ c = \sum_a \lambda_a \left( 1 - \Gamma_{0a} \right) |\delta \phi_{\bf k}|, $$
and then obtain
	$$ \sum_a n_a T_a s_a \ge \left[ \sum_a \lambda_a \left( 1 - \Gamma_{0a} \right) |\delta \phi_{\bf k}|\right ]^2 \bigg\slash \sum_c \lambda_c \Gamma_{0c}. $$
As a result of this inequality, we conclude from Eq.\,(\ref{H}) that $ H \ge \lang L |\delta \phi_{\bf k} |^2 \rang $ with 
	$$ L(l) = \left( \sum_a \lambda_a \right) \left( \sum_b \lambda_b (1-\Gamma_{0b}) \right) \bigg\slash \left( \sum_c \lambda_c \Gamma_{0c} \right). $$
Similarly, by instead choosing $c = |k_\perp \delta A_{\|\bf k} |/\mu_0$ and
	$$ p_a = \frac{n_a |e_a|}{k_\perp} \sqrt{ \frac{T_a \Gamma_{0a}}{m_a}}, $$
we find
	$$ \sum_a n_a T_a s_a \ge \frac{|k_\perp \delta A_{\|\bf k} |^2}{\mu_0} \bigg\slash \sum_a \frac{\beta_a \Gamma_{0a}}{2 b_a}, $$
where $\beta_a = 2 \mu_0 n_a T_a / B^2$. Because the gyroradius of the electrons is usually much smaller than that of any ion species and $\Gamma_{a0} = \Gamma_0(b_a)$ is a decreasing function of particle mass, only the electrons need to be kept in the sum over species, and we conclude that $H$ is bounded from below by
	$$ H({\bf k},t) \ge \lang \frac{|k_\perp \delta A_{\|\bf k} |^2}{\mu_0} \left( 1 + \frac{2 b_e}{\beta_e \Gamma_{0e}} \right) \rang
	= \frac{ne^2}{m_e} \lang K |\delta A_{\|\bf k} |^2 \rang, $$
with
	$$ K(l) = \frac{2 b_e}{\beta_e}	\left( 1 + \frac{2 b_e}{\beta_e \Gamma_{0e}} \right). $$
	
We are now ready to apply our basic upper bound (\ref{bound on linear growth}), where we use Eq.~(\ref{D}) and
	$$ H \ge \lang n_a T_a s_a \rang^{1/2} \lang L |\delta \phi_{\bf k} |^2 \rang^{1/2}, $$
	$$ H \ge \lang n_a T_a s_a \rang^{1/2} \lang \frac{ne^2}{m_e} K |\delta A_{\|\bf k} |^2 \rang^{1/2}, $$
to conclude that
	$$ \gamma \le \sum_a |\omega_{\ast a}| 
	\sqrt{\frac{\lang \lambda_a M(\eta_a, b_a) |\delta \phi_{\bf k}|^2 \rang}{\lang L |\delta \phi_{\bf k}|^2 \rang}} 
	+ |\omega_{\ast e}| 
	\sqrt{\frac{\lang N(\eta_e, b_e) |\delta A_{\| \bf k} |^2 \rang}{\lang K |\delta A_{\| \bf k} |^2\rang}} 
	$$
where the contribution from ions to the electromagnetic term in $D$ has been neglected, being a factor of order $m_e/m_i$ smaller than the electron contribution. Since $L$ is an increasing function of the quantities $b_a$, which are all proportional to $(k_\perp / B)^2$, the first term on the right is maximized if $|\delta \phi_{\bf k}(l)|^2$ is chosen to be delta function in the point $l_0$ where the function $k_\perp(l) / B(l)$ attains its minimum. Similarly, the second term is maximized by choosing $| \delta A_{\| \bf k}(l) |^2 \propto \delta(l-l_1)$ where $l_1$ is the point where $K(l)/N(l)$ is minimized. We thus arrive at the result	
	 \bn \gamma({\bf k}) \le \gamma_{\rm bound} ({\bf k}) = \sum_a |\omega_{\ast a}| 
	\sqrt{\frac{\lambda_a M(\eta_a, b_a(l_0))}{L(l_0)}} 
	+ |\omega_{\ast e}| 
	\sqrt{\frac{ N(\eta_e, b_e(l_1))}{K(l_1) }}, \label{most general bound}
	\en
Apart from the neglect of terms of order $m_e/m_i$ and fluctuations in the magnetic-field strength, $\delta B_\|$, this upper bound on the growth rate is completely general and applies to any local gyrokinetic instability. It applies to ion- and electron-temperature-gradient modes, kinetic and resistive ballooning modes, trapped-ion and trapped-electron modes, and microtearing modes, as well as to the so-called universal and ubiquitous instabilities. 

A particularly simple and important case is that of a hydrogen plasma without other ions and $k_\perp \rho_e \ll 1$. Noting that $\omega_{\ast i} = - \tau \omega_{\ast e}$ and using the asymptotic forms (\ref{M}) and (\ref{N}), we find
	\bn \frac{\gamma}{|\omega_{\ast e}|} \le \sqrt{\frac{\tau(\Gamma_{0i} + \tau)}{(1 + \tau) (1 - \Gamma_{0i})} }
	\left( \sqrt{\tau M(\eta_i, b_i)} + \sqrt{1 + \frac{3 \eta_e^2}{2}} \right) 
	+ \beta_e \sqrt{\frac{ 1 + 2 \eta_a + 7 \eta_e^2/2}{2 b_e \left(\beta_e + 2 b_e \right)}}, 
	\label{hydrogen}
	\en
where the first term on the right is evaluated at $l=l_0$ and the second one (which is proportional to $\beta_e$) at $l=l_1$. Both terms give an upper bound on $\gamma$ that remains finite in the short-wavelength limit since $\omega_{\ast e}$ is proportional to $k_\alpha$ and 
	$$ 1 - \Gamma_{0i} \simeq b_i = (k_\perp \rho_i)^2, $$
in the limit $b_i \ll 1$. Thus, as long as $k_\perp \rho_e \ll 1$, the growth rate is subject to a bound equal to 
	\bn \gamma < C_0 \left( 1 + \tau^{-1/2} \right) \frac{v_{Ti}}{L_\perp} 
	+ \frac{C_1 \beta_e}{\sqrt{\beta_e + 2 b_e}} \frac{v_{T_e}}{L_\perp}, 
	\label{em scaling}
	\en
where $C_0$ and $C_1$ are numbers of order unity, $v_{Ti}$ denotes the ion thermal speed, and $L_\perp$ the length scale of the equilibrium density and temperature gradients. In the opposite limit, $k_\perp \rho_e \gg 1$, the term proportional to $\beta_e$ can be neglected and we instead obtain
		\bn {\gamma} \le \frac{\tau {|\omega_{\ast e}|}}{1 + \tau} \sqrt{\frac{1-\eta_e + 5 \eta_e^2/4}{2 \pi b_e(l_0)}} = \frac{C_2 v_{Te}}{(1 + \tau^{-1}) L_\perp},
		\en
where $v_{Te}$ denotes the electron thermal speed and $C_2$ is a number of order unity.

\section{Optimal bounds}

The bounds (\ref{ae}) and (\ref{most general bound}) are not optimal and can be improved. In this section, we derive the best possible bound, in a sense that will be made precise, for the simplest case of a hydrogen plasma with adiabatic electrons. If $\varphi = e \delta \phi_{\bf k} / T_i$ and $g = g_{i\bf k}$, we have 
	$$ \varphi = \frac{1}{n (1+\tau)} \int g J_{0} d^3v, $$
	$$ H = nT_i \lang \frac{1}{n} \int \frac{|g|^2}{F_{i0}} d^3v - (1+\tau) |\varphi|^2 \rang, $$
	$$ D = \frac{\eta_i \omega_{\ast  i} T_i}{2i} \lang \int \left(\varphi^\ast g - \varphi g^\ast\right) x^2 J_{0i} d^3v \rang, $$
where $x^2 = m_i v^2 / 2 T_i$. $D$ and $H$ are thus quadratic functionals of $g$, and the challenge is to maximise the ratio $D[g]/H[g]$ over all such functions.

In order to do so, we first note that $D$ and $\varphi$ only depend on two moments of $g$, namely, 
	$$ K_j[g] = \frac{1}{n} \int g x^{2j} J_{0i} d^3v, $$
where $j = 0$ or 1. We can therefore begin by minimising $H[g]$ over all functions with given values of these two moments. Using Lagrange multipliers, $c_0$ and $c_1$, we are thus led to minimise the functional
	$$ H[g] - 2c_0 K_0[g] - 2c_1 K_1[g], $$
which gives
	\bn g = \left( c_0 + c_1 x^2 \right) J_{0i} F_{i0}. 
	\label{g}
	\en
We have thus reduced our problem to that of finding the maximum value of $D/H$ expressed as a ratio of two quadratic forms in the coefficients $c_j$. 

If we write 
	$$ G_j(b_i) = \frac{1}{n} \int F_{i0} x^{2j} J_{0i}^2 d^3v, $$
so that 
	$$ G_0(b_i) = \Gamma_0(b_i), $$
	$$ G_1(b_i) = \left( \frac{3}{2} - b_i \right) \Gamma_0(b_i) + b_i \Gamma_1(b_i), $$
	$$ G_2(b_i) = \left( \frac{15}{4} - 5b_i  + 2 b_i^2 \right) \Gamma_0(b_i) 
	+ \left( 4 - 2 b_i \right) b_i \Gamma_1(b_i), $$
then
	$$ D = \frac{n T_i G(b_i) }{2i(1+\tau)} \left( c_0^\ast c_1 - c_0 c_1^\ast \right), $$
where 
	$$G(b) = G_0(b_i) G_2(b_i) - G_1^2(b_i)
	= \left( \frac{3}{2} - 2 b_i + b_i^2 \right) \Gamma_0^2(b_i) + b_i \Gamma_0 (b_i) \Gamma_1(b_i)
	- b_i^2 \Gamma_1^2 (b_i^2), $$
and
	$$ H = n T_i \left[ G_0 \left( 1 - \frac{G_0}{1+\tau} \right) c_0 c_0^\ast  
	+ G_1 \left( 1 - \frac{G_0}{1+\tau} \right) \left( c_0^\ast c_1 + c_0 c_1^\ast \right) 
	+ \left( G_2 - \frac{G_1^2}{1+\tau} \right) c_0 c_0^\ast  \right]. $$

In order to maximise the ratio and calculate
	$$ \hat \gamma = \max_{c_0, c_1} \left( \frac{D}{H} \right) , $$ 
we consider the variations
	$$ \delta D = \frac{n T_i G }{2i(1+\tau)} \left( c_1 \delta c_0^\ast - c_0 \delta c_1^\ast \right) + \rm c.c., $$
	\begin{eqnarray*} 
	\delta H &=& n T_i \left[ G_0 \left( 1 - \frac{G_0}{1+\tau} \right) c_0 \delta c_0^\ast 
	+ G_1 \left( 1 - \frac{G_0}{1+\tau} \right) \left( c_1 \delta c_0^\ast  + c_0 \delta c_1^\ast \right) 
		+ \left( G_2 - \frac{G_1^2}{1+\tau} \right) c_1 \delta c_1^\ast  \right] \\
		& & + \ \rm c.c., 
	\end{eqnarray*}
where $\rm c.c.$ stands for the complex conjugate, and we note that the maximum is reached when
	\bn \delta D = \hat \gamma \delta H, 
	\label{eigenvalue problem}
	\en
which gives a system of equations
	\begin{eqnarray*} \frac{2i\hat \gamma}{\eta_i \omega_{\ast i}}  \left[ \begin{array}{cc} G_0 \left( 1 + \tau - G_0 \right) & G_1 \left( 1 + \tau - G_0 \right) \\
	 G_1 \left( 1 + \tau - G_0 \right) & G_2(1 + \tau) - G_1^2 	\end{array} \right]
	\left[ \begin{array}{c} c_0 \\ c_1 \end{array} \right] = G \left[ \begin{array}{c} c_1 \\ c_0 \end{array} \right],
	\end{eqnarray*}
which has non-zero solutions if
	\bn \hat \gamma = \frac{|\eta_i \omega_{\ast i}|}{2} \sqrt{\frac{G(b_i)}{(1+\tau) [1 + \tau - G_0(b_i)]}}. 
	\label{optimal ae bound}
	\en
	
This is the ``optimal'' bound on the growth rate that can be obtained within our formalism in the sense that no lower bound is possible. Indeed, growth of the free energy at this rate is realised if no collisions are present and the distribution function is chosen as dictated by Eq.\,(\ref{g}) with $c_0$ and $c_1$ satisfying the eigenvalue problem (\ref{eigenvalue problem}). The bound (\ref{optimal ae bound}) is shown in Fig.~1 and is lower than our previous result (\ref{ae}) by a factor of 2 and $\sqrt{5}$ in the limits of long and short wavelengths, respectively, 
	\begin{eqnarray} \hat \gamma \rightarrow \left\{ \begin{array}{c l} \frac{|\eta_i \omega_{\ast i}|}{2} \sqrt{\frac{3}{2 \tau (1+\tau)}} , & b_i \ll 1
	\\ \frac{| \eta_i \omega_{\ast i}|}{(1+\tau) \sqrt{8 \pi b_i}}, & b_i \gg 1. \end{array} \right. 
	\end{eqnarray}
	
\section{Bounds on nonlinear growth}

Our most general bound (\ref{most general bound}) is not optimal and will be improved substantially in our next publication, but its most important implication follows already from this crude form. The right-hand side is a bounded function of the mode numbers $(k_\psi, k_\alpha)$, and the linear growth rate can therefore never exceed the maximum
	\bn \gamma_{\rm max} = \sup_{\bf k} \; \gamma_{\rm bound} ({\bf k}). 
	\label{gamma-max}
	\en
As we shall now see, this conclusion also holds for nonlinear growth. 

Consider the evolution of a set of fluctuations governed by the gyrokinetic system of equations starting from some arbitrary initial condition, specified by the distribution functions $\delta F_a$ of all species at $t=0$. According to Eq.\,(\ref{entropy balance}) the instantaneous growth of the total free energy,
	$$ H_{\rm tot}(t) = \sum_{\bf k} H({\bf k},t) $$
is bounded by 
	$$ \frac{d H_{\rm tot}}{dt} \le 2  \sum_{\bf k} D({\bf k},t), $$
where each term is subject to the bound
	$$ D({\bf k},t) \le \gamma_{\rm bound}({\bf k}) H({\bf k},t). 
	$$
The growth rate of the total free energy is therefore limited by twice the maximum linear growth
	$$ \frac{d \ln H_{\rm tot}}{dt} \le 2 \gamma_{\rm max}. $$
This bound holds for fluctuations of arbitrary amplitude within the gyrokinetic formalism. In particular, it must hold in any gyrokinetic simulation of turbulence. 

Moreover, if collisions are absent, then instantaneous growth of the free energy is possible at any positive rate up to the ``optimal'' one, which for the particularly simple case of adiabatic electrons was derived in the previous subsection. To see this, suppose the bounds on the right-hand side of Eq.\,(\ref{gamma-max}) are chosen optimally in the sense that
	$$ \gamma_{\rm bound} ({\bf k}) = \sup_{g} \; \frac{D[g,{\bf k}]}{H[g,{\bf k}]}, $$
where $D$ and $H$ are considered to be quadratic functionals of the distribution functions $g = \{ g_a \}$ of all species. This means, then, that there is a choice of wave number and initial data such that the free energy grows at a rate arbitrarily close to $2 \gamma_{\rm max}$. Conversely, there is a similar limit on the rate at which the free energy can decay in the absence of collisions, 
	$$ \frac{d \ln H_{\rm tot}}{dt} \ge - 2 \gamma_{\rm max}, $$
as follows from the observation that $D$ is odd in the wave number $\bf k$ whereas $H$ is even. The transformation ${\bf k} \rightarrow - {\bf k}$ thus changes the sign of the ratio ${D[g,{\bf k}]}/{H[g,{\bf k}]}$. Any upper bound on the latter therefore automatically implies a similar lower bound when collisions are absent.

\section{Conclusions}

As we have seen, it is possible to derive rigorous upper bounds on the growth rate of linear instabilities and on the nonlinear growth of free energy in gyrokinetics. Unlike most other results in the field, these bounds are universal and hold in plasmas with any number of particle species regardless of collisionality and magnetic-field geometry. For simplicity, we have taken the plasma pressure (beta) to be sufficiently small that fluctuations in the magnetic-field strength can be neglected, $\delta B_\| = 0$, but this restriction will be removed in Part II in the present series of papers. 

In the case of a plasma with a single kinetic ion species and ``adiabatic'' electrons, the bound is given by Eq.~(\ref{optimal ae bound}) and is of order
	$$ \gamma_{\rm max}  \sim \frac{k_\perp \rho_i}{\sqrt{\tau (1+\tau)}} \cdot \frac{ v_{Ti}}{L_\perp }$$
for $k_\perp \rho_i < 1$ and 
	$$ \gamma_{\rm max} \sim \frac{v_{Ti}}{(1 + \tau) L_\perp} $$
for shorter wavelengths. The dependence on the parameter $\tau = T_i/T_e$ reflects a well-known unfavourable dependence of the ITG growth rate on electron temperature. 

The bound (\ref{most general bound}) we found on instabilities with kinetic electrons is less restrictive and remains finite in the limit $k_\perp \rho_i \rightarrow 0$. It is a sum of two distinct contributions: an electrostatic term and an electromagnetic term that vanishes if $\beta_e \rightarrow 0$. As we shall see in the next publication of this series, this result is not qualitatively affected by the inclusion of parallel magnetic fluctuations. 

Actual microinstability growth rates must lie below these bounds. For instance, toroidal ITG modes with adiabatic electrons and $k_\perp \rho_i \ll 1$ have growth rates 
	$$ \gamma \sim \sqrt{\frac{\eta_i \omega_{\ast i} \omega_{di}}{\tau} } \sim \frac{k_\perp \rho_i}{\sqrt{\tau}} \cdot \frac{ v_{Ti}}{\sqrt{R L_\perp} }$$
in the strongly-driven limit \citep{Bigliari-1989,Romanelli-1989,Plunk-2014,Zocco-2018}, and trapped-ion modes have a similar growth rate \cite{Bigliari-1989}. Here $R$ denotes the radius of curvature of the magnetic field, so that $\omega_{di} \sim (k_\perp \rho_i) v_{Ti} / R$. Due to the assumption $| \omega_{di} /\omega_{\ast i} |\sim L_\perp / R \ll 1$ (corresponding to strong instability drive) made in the derivation of this estimate, the growth rate is smaller than our upper bound. Similarly, in the theory of kinetic ballooning modes, the assumption $L_\perp / R \ll 1$ leads to growth rates of order \citep{Tang-1980,Aleynikova-2018}
		$$  \gamma \sim \frac{\sqrt{\omega_{di} \left[ (1 + \eta_i) \omega_{\ast i} - (1 + \eta_e) \omega_{\ast e} \right]}}{k_\perp \rho_i}. $$
This growth rate never exceeds our bound (\ref{hydrogen}) and scales as our estimate (\ref{em scaling}). In less strongly driven cases, the growth rate is lower. 

Although all our results are quite general, they do not encompass all instabilities of interest. Kink modes and tearing modes sometimes need a gyrokinetic treatment in a thin layer around a resonant magnetic surface, where magnetic reconnection may occur, but take their energy from the exterior region and depend on the overall plasma current profile \citep{Hazeltine-1975,Drake-1977}. Such instabilities cannot adequately be described in the geometry of a magnetic flux tube \citep{Connor-2014,Connor-2019} and are not subject to the bounds derived in the present paper. Mathematically, they are not covered by our treatment since the solution of the gyrokinetic equation involves matching to the exterior region, whose destabilising influence is usually described by a parameter $\Delta'$, making these modes non-local in nature. However, microtearing modes which are driven by local gradients are subject to our bound (\ref{most general bound}) on electromagnetic instabilities. 
\\ \\
This work was partly supported by a grant from the Simons Foundation (560651, PH). 

\section*{Appendix: a quadratic minimsation problem}

Consider the problem of minimising
	$$ f({\bf x}) = \sum_a q_a x_a^2 $$
where ${\bf x} = (x_1, x_2, \cdots)$ subject to the constraint 
	$$ \sum_a p_a x_a  \ge c, $$ 
where $q_a$ and $p_a$ are positive real numbers. This problem is not difficult to solve by considering the function
	$$ F({\bf x},\lambda) = f({\bf x}) - \lambda \left( \sum_a p_a x_a - c \right), $$
where $\lambda$ is a Lagrange multiplier. The conditions
	$$ \frac{\partial F}{\partial x_a} = \frac{\partial F}{\partial \lambda} = 0 $$
lead to 
		$$ x_a = \frac{\lambda p_a}{2q_a}, $$
		$$ \lambda = 2 c \bigg\slash \sum_a \frac{p_a^2}{q_a} , $$
and 
	$$ \min_{\bf x} f({\bf x}) = c^2 \bigg\slash \sum_a \frac{p_a^2}{q_a}. $$

\bibliographystyle{jpp}
\bibliography{modes}

\begin{thebibliography}{34}
\expandafter\ifx\csname natexlab\endcsname\relax\def\natexlab#1{#1}\fi
\def\au#1{#1} \def\ed#1{#1} \def\yr#1{#1}\def\at#1{#1}\def\jt#1{\textit{#1}}
  \def\bt#1{#1}\def\bvol#1{\textbf{#1}} \def\vol#1{#1} \def\pg#1{#1}
  \def\publ#1{#1}\def\arxiv#1{#1}\def\org#1{#1}\def\st#1{\textit{#1}}

\bibitem[Aleynikova {\em et~al.\/}({2018})Aleynikova, Zocco, Xanthopoulos,
  Helander \& N{\"u}hrenberg]{Aleynikova-2018}
{\sc \au{Aleynikova, K.}, \au{Zocco, A.}, \au{Xanthopoulos, P.}, \au{Helander,
  P.} \& \au{N{\"u}hrenberg, C.}} \yr{{2018}}  \at{Kinetic ballooning modes in
  tokamaks and stellarators}.  \jt{{J. Plasma Phys.}}  \bvol{{84}}~({6}),
  \pg{{745840602}}.

\bibitem[Antonsen \& Lane({1980})]{Antonsen-1980}
{\sc \au{Antonsen, T.M.} \& \au{Lane, B.}} \yr{{1980}}  \at{Kinetic-equations
  for low-frequency instabilities in inhomogeneous plasmas}.  \jt{{Phys.
  Fluids}}  \bvol{{23}}~({6}),  \pg{{1205--1214}}.

\bibitem[Banon~Navarro {\em et~al.\/}(2011)Banon~Navarro, Morel, Albrecht-Marc,
  Carati, Merz, Goerler \& Jenko]{Banon-2011}
{\sc \au{Banon~Navarro, A.}, \au{Morel, P.}, \au{Albrecht-Marc, M.},
  \au{Carati, D.}, \au{Merz, F.}, \au{Goerler, T.} \& \au{Jenko, F.}} \yr{2011}
   \at{Free energy cascade in gyrokinetic turbulence}.  \jt{Phys. Rev. Lett.}
  \bvol{106},  \pg{055001}.

\bibitem[Bigliari {\em et~al.\/}({1989})Bigliari, Diamond \&
  Rosenbluth]{Bigliari-1989}
{\sc \au{Bigliari, H.}, \au{Diamond, P.H.} \& \au{Rosenbluth, M.N.}}
  \yr{{1989}}  \at{Toroidal ion-pressure-gradient-driven drift instabilities
  and transport revisited}.  \jt{{Phys. Fluids B}}  \bvol{{1}}~({1}),
  \pg{{109--118}}.

\bibitem[Brizard({1994})]{Brizard-1994}
{\sc \au{Brizard, A.J.}} \yr{{1994}}  \at{{Quadratic free energy for the
  linearized gyrokinetic Vlasov-Maxwell equations}}.  \jt{{Phys. Plasmas}}
  \bvol{{1}}~({8}),  \pg{{2473--2479}}.

\bibitem[Brizard {\em et~al.\/}({1991})Brizard, Fowler, Hua \&
  Morrison]{Brizard-1991}
{\sc \au{Brizard, A.J.}, \au{Fowler, T.K.}, \au{Hua, D.} \& \au{Morrison,
  P.J.}} \yr{{1991}}  \at{{Thermodynamic constraints applied to tokamaks}}.
  \bt{In {\em {Comments on Plasma Physics and Controlled Fusion}\/}}, ,
  \vol{vol.~{14}},  \pg{pp. {263--273}}.  \publ{Gordon and Breach, Science
  Publishers S.A.}

\bibitem[Brizard \& Hahm({2007})]{Brizard-2007}
{\sc \au{Brizard, A.J.} \& \au{Hahm, T.S.}} \yr{{2007}}  \at{{Foundations of
  nonlinear gyrokinetic theory}}.  \jt{{Rev. Mod. Phys.}}  \bvol{{79}}~({2}),
  \pg{{421--468}}.

\bibitem[Catto({1978})]{Catto-1978-a}
{\sc \au{Catto, P.J.}} \yr{{1978}}  \at{Linearized gyro-kinetics}.  \jt{{Plasma
  Phys.}}  \bvol{{20}}~({7}),  \pg{{719--722}}.

\bibitem[Catto({2019})]{Catto-2019}
{\sc \au{Catto, P.J.}} \yr{{2019}}  \at{{Practical gyrokinetics}}.  \jt{{J.
  Plasma Phys.}}  \bvol{{85}}~({3}),  \pg{{925850301}}.

\bibitem[Catto {\em et~al.\/}({1981})Catto, Tang \& Baldwin]{Catto-1981}
{\sc \au{Catto, P.J.}, \au{Tang, W.M.} \& \au{Baldwin, D.E.}} \yr{{1981}}
  \at{Generalized gyrokinetics}.  \jt{{Plasma Phys.}}  \bvol{{23}}~({7}),
  \pg{{639--650}}.

\bibitem[Connor {\em et~al.\/}({2019})Connor, Ham, Hastie \&
  Zocco]{Connor-2019}
{\sc \au{Connor, J.W.}, \au{Ham, C.J.}, \au{Hastie, R.J.} \& \au{Zocco, A.}}
  \yr{{2019}}  \at{{Ion Landau damping and drift tearing modes}}.  \jt{{J.
  Plasma Phys.}}  \bvol{{85}}~({2}),  \pg{{905850204}}.

\bibitem[Connor {\em et~al.\/}({2014})Connor, Hastie, Pusztai, Catto \&
  Barnes]{Connor-2014}
{\sc \au{Connor, J.W.}, \au{Hastie, R.J.}, \au{Pusztai, I.}, \au{Catto, P.J.}
  \& \au{Barnes, M.}} \yr{{2014}}  \at{{High-m kink/tearing modes in
  cylindrical geometry}}.  \jt{{Plasma Phys. Contr. Fusion}}
  \bvol{{56}}~({12}),  \pg{{125006}}.

\bibitem[Drake \& Lee(1977)]{Drake-1977}
{\sc \au{Drake, J.F.} \& \au{Lee, Z.C.}} \yr{1977}  \at{Kinetic theory of
  tearing instabilities}.  \jt{Phys. Fluids}  \bvol{20},  \pg{1341}.

\bibitem[Fowler({1964})]{Fowler-1964}
{\sc \au{Fowler, T.K.}} \yr{{1964}}  \at{{Bounds on Plasma Instability Growth
  Rates}}.  \jt{{Phys. Fluids}}  \bvol{{7}}~({2}),  \pg{{249--256}}.

\bibitem[Fowler(1968)]{Fowler-1968}
{\sc \au{Fowler, T.K.}} \yr{1968}  \at{Thermodynamics of unstable plasmas}.
  \bt{In {\em Advances in plasma physics\/} (ed. \ed{A.~Simon \& W.B.
  Thompson})}, ,  \vol{vol.~1},  \pg{p. 201}.  \publ{New York: John Wiley and
  Sons, Inc.}

\bibitem[Frieman \& Chen({1982})]{Frieman-1982}
{\sc \au{Frieman, E.A.} \& \au{Chen, L.}} \yr{{1982}}  \at{Non-linear
  gyrokinetic equations for low-frequency electromagnetic-waves in general
  plasma equilibria}.  \jt{{Phys. Fluids}}  \bvol{{25}}~({3}),
  \pg{{502--508}}.

\bibitem[Garbet {\em et~al.\/}({2005})Garbet, Dubuit, Asp, Sarazin, Bourdelle,
  Ghendrih \& Hoang]{Garbet-2005}
{\sc \au{Garbet, X.}, \au{Dubuit, N.}, \au{Asp, E.}, \au{Sarazin, Y.},
  \au{Bourdelle, C.}, \au{Ghendrih, P.} \& \au{Hoang, G.T.}} \yr{{2005}}
  \at{{Turbulent fluxes and entropy production rate}}.  \jt{{Phys. Plasmas}}
  \bvol{{12}},  \pg{{082511}}.

\bibitem[Garbet {\em et~al.\/}({2010})Garbet, Idomura, Villard \&
  Watanabe]{Garbet-2010}
{\sc \au{Garbet, X.}, \au{Idomura, Y.}, \au{Villard, L.} \& \au{Watanabe,
  T.H.}} \yr{{2010}}  \at{{Gyrokinetic simulations of turbulent transport}}.
  \jt{{Nucl. Fusion}}  \bvol{{50}}~({4}),  \pg{{043002}}.

\bibitem[Hatch {\em et~al.\/}(2016)Hatch, Jenko, Navarro, Bratanov, Terry \&
  Pueschel]{Hatch-2016}
{\sc \au{Hatch, D~R}, \au{Jenko, F}, \au{Navarro, A~Ba{\~{n}}{\'{o}}n},
  \au{Bratanov, V}, \au{Terry, P~W} \& \au{Pueschel, M~J}} \yr{2016}
  \at{Linear signatures in nonlinear gyrokinetics: {I}nterpreting turbulence
  with pseudospectra}.  \jt{New Journal of Physics}  \bvol{18}~(7),
  \pg{075018}.

\bibitem[Hazeltine {\em et~al.\/}({1975})Hazeltine, Dobrott \&
  Wang]{Hazeltine-1975}
{\sc \au{Hazeltine, R.D.}, \au{Dobrott, D.} \& \au{Wang, T.S.}} \yr{{1975}}
  \at{Kinetic theory of tearing instability}.  \jt{{Phys. Fluids}}
  \bvol{{18}}~({12}),  \pg{{1778--1786}}.

\bibitem[Helander \& Plunk({2021})]{Helander-Plunk-2021}
{\sc \au{Helander, P.} \& \au{Plunk, G.G.}} \yr{{2021}}  \at{{Upper Bounds on
  Gyrokinetic Instabilities in Magnetized Plasmas}}.  \jt{{Phys. Rev. Lett.}}
  \bvol{{127}},  \pg{{155001}}.

\bibitem[Kotschenreuther {\em et~al.\/}({1995})Kotschenreuther, Rewoldt \&
  Tang]{Kotschenreuther-1995}
{\sc \au{Kotschenreuther, M.}, \au{Rewoldt, G.} \& \au{Tang, W.M.}} \yr{{1995}}
   \at{Comparison of initial-value and eigenvalue codes for kinetic toroidal
  plasma instabilities}.  \jt{{Comp. Phys. Comm.}}  \bvol{{88}}~({2-3}),
  \pg{{128--140}}.

\bibitem[Krommes({2012})]{Krommes-2012}
{\sc \au{Krommes, J.A.}} \yr{{2012}}  \at{The gyrokinetic description of
  microturbulence in magnetized plasmas}.  \jt{{Ann. Rev. Fluid Mech.}}
  \bvol{{44}},  \pg{{175--201}}.

\bibitem[Krommes \& Hu({1993})]{Krommes-1993}
{\sc \au{Krommes, John~A.} \& \au{Hu, Genze}} \yr{{1993}}  \at{General theory
  of onsager symmetries for perturbations of equilibrium and nonequilibrium
  steady states}.  \jt{{Phys. Fluids B}}  \bvol{{5}},  \pg{{3908}}.

\bibitem[Plunk {\em et~al.\/}({2014})Plunk, Helander, Xanthopoulos \&
  Connor]{Plunk-2014}
{\sc \au{Plunk, G.G.}, \au{Helander, P.}, \au{Xanthopoulos, P.} \& \au{Connor,
  J.W.}} \yr{{2014}}  \at{{Collisionless microinstabilities in stellarators.
  III. The ion-temperature-gradient mode}}.  \jt{{Phys. Plasmas}}
  \bvol{{21}}~({3}),  \pg{{032112}}.

\bibitem[Romanelli({1989})]{Romanelli-1989}
{\sc \au{Romanelli, F.}} \yr{{1989}}  \at{Ion temperature‐gradient‐driven
  modes and anomalous ion transport in tokamaks}.  \jt{{Phys. Fluids B}}
  \bvol{{1}}~({5}),  \pg{{1018--1025}}.

\bibitem[Rutherford \& Frieman({1968})]{Rutherford-1968-a}
{\sc \au{Rutherford, P.H.} \& \au{Frieman, E.A.}} \yr{{1968}}  \at{Drift
  instabilities in general magnetic field configurations}.  \jt{{Phys. Fluids}}
   \bvol{{11}}~({3}),  \pg{{569--585}}.

\bibitem[Schekochihin {\em et~al.\/}({2009})Schekochihin, Cowley, Dorland,
  Hammett, Howes, Quataert \& Tatsuno]{Schekochihin-2009}
{\sc \au{Schekochihin, A.A.}, \au{Cowley, S.C.}, \au{Dorland, W.}, \au{Hammett,
  G.W.}, \au{Howes, G.G.}, \au{Quataert, E.} \& \au{Tatsuno, T.}} \yr{{2009}}
  \at{Astrophysical gyrokinetics: Kinetic and fluid turbulent cascades in
  magnetized weakly collisional plasmas}.  \jt{{Astrophys. J. Suppl. S.}}
  \bvol{{182}}~({1}),  \pg{{310--377}}.

\bibitem[Stoltzfus-Dueck \& Scott({2017})]{Stoltzfus-Dueck-2017}
{\sc \au{Stoltzfus-Dueck, T.} \& \au{Scott, B.}} \yr{{2017}}  \at{Momentum flux
  parasitic to free-energy transfer}.  \jt{{Nucl. Fusion.}}  \bvol{{57}}~({8}),
   \pg{{086036}}.

\bibitem[Sugama {\em et~al.\/}({1996})Sugama, Okamoto, Horton \&
  Wakatani]{Sugama-1996}
{\sc \au{Sugama, H.}, \au{Okamoto, M.}, \au{Horton, W.} \& \au{Wakatani, M.}}
  \yr{{1996}}  \at{{Transport processes and entropy production in toroidal
  plasmas with gyrokinetic electromagnetic turbulence}}.  \jt{{Phys. Plasmas}}
  \bvol{{3}}~({6}),  \pg{{2379--2394}}.

\bibitem[Tang {\em et~al.\/}(1980)Tang, Connor \& Hastie]{Tang-1980}
{\sc \au{Tang, W.M.}, \au{Connor, J.W.} \& \au{Hastie, R.J.}} \yr{1980}
  \at{Kinetic-ballooning-mode theory in general geometry}.  \jt{Nucl. Fusion}
  \bvol{20}~(11),  \pg{1439}.

\bibitem[Tatsuno {\em et~al.\/}({2009})Tatsuno, Dorland, Schekochihin, Plunk,
  Barnes, Cowley \& Howes]{Tatsuno-2009}
{\sc \au{Tatsuno, T.}, \au{Dorland, W.}, \au{Schekochihin, A.A.}, \au{Plunk,
  G.G.}, \au{Barnes, M.}, \au{Cowley, S.C.} \& \au{Howes, G.G.}} \yr{{2009}}
  \at{Nonlinear phase mixing and phase-space cascade of entropy in gyrokinetic
  plasma turbulence}.  \jt{{Phys. Rev. Lett.}}  \bvol{{103}},  \pg{{015003}}.

\bibitem[Taylor \& Hastie({1968})]{Taylor-1968}
{\sc \au{Taylor, J.B.} \& \au{Hastie, R.J.}} \yr{{1968}}  \at{Stability of
  general plasma equilibria {.I. F}ormal theory}.  \jt{{Plasma Phys.}}
  \bvol{{10}}~({5}),  \pg{{479--494}}.

\bibitem[Zocco {\em et~al.\/}({2018})Zocco, Plunk, Xanthopoulos \&
  Helander]{Zocco-2018}
{\sc \au{Zocco, A.}, \au{Plunk, G.G.}, \au{Xanthopoulos, P.} \& \au{Helander,
  P.}} \yr{{2018}}  \at{{Threshold for the destabilisation of the
  ion-temperature-gradient mode in magnetically confined toroidal plasmas}}.
  \jt{{J. Plasma Phys.}}  \bvol{{84}},  \pg{{715840101}}.

\end{thebibliography}

\end{document}